\documentclass[12pt,english,floatfix,superscriptaddress,aps,prd,preprint]{revtex4}
\usepackage[utf8]{inputenc}
\usepackage{amsmath}
\usepackage{amssymb}
\usepackage{amsbsy}
\usepackage{amsfonts}
\usepackage{amsopn}
\usepackage{amstext}
\usepackage{graphicx}
\usepackage{amssymb}
\usepackage{amsfonts}
\usepackage{amsmath}
\usepackage{amsmath,amsthm,amsfonts,amssymb}
\usepackage[mathcal]{eucal}
\usepackage{mathrsfs}
\usepackage{graphicx}
\usepackage[english]{babel}
\usepackage{color}
\usepackage{esint}
\usepackage[dvips]{epsfig}
\usepackage[dvips]{graphicx}
\usepackage{float}
\usepackage{units}
\usepackage{textcomp}

\usepackage{hyperref}
\usepackage{slashed}

\newcommand{\ie}{\begin{equation}}
\newcommand{\fe}{\end{equation}}
\newcommand{\se}{\begin{eqnarray}}
\newcommand{\ff}{\end{eqnarray}}

\usepackage{xcolor}
\usepackage{soul}

\begin{document}

\title{  Information and thermodynamic properties of a non-Hermitian particle ensemble   }


\author{F. C. E. Lima}
\email{cleiton.estevao@fisica.ufc.br}
\affiliation{Universidade Federal do Cear\'a (UFC), Departamento de F\'isica,\\ Campus do Pici, Fortaleza - CE, C.P. 6030, 60455-760 - Brazil.}


\author{A. R. P. Moreira}
\email{allan.moreira@fisica.ufc.br}
\affiliation{Universidade Federal do Cear\'a (UFC), Departamento de F\'isica,\\ Campus do Pici, Fortaleza - CE, C.P. 6030, 60455-760 - Brazil.}


\author{C. A. S. Almeida}
\email{carlos@fisica.ufc.br}
\affiliation{Universidade Federal do Cear\'a (UFC), Departamento de F\'isica,\\ Campus do Pici, Fortaleza - CE, C.P. 6030, 60455-760 - Brazil.}

\date{\today}

\begin{abstract}
 In the context of non-relativistic quantum mechanics, we investigated Shannon's entropy of a non-Hermitian system to understand how this quantity is modified with the cyclotron frequency. Subsequently, we turn our attention to the construction of an ensemble of these spinless particles in the presence of a uniform magnetic field. Then, we study the thermodynamic properties of the model. Finally, we show how Shannon's entropy and thermodynamic properties are modified with the action of the magnetic field.
 \end{abstract}

\maketitle

\section{Introduction}

In the initial course of quantum mechanics, we learned that the Hamiltonian operator must be Hermitian \cite{Griffiths, Sakurai}. This premise is extremely important for us to have the guarantee of obtaining an energy spectrum with real eigenvalues \cite{Griffiths}. As a consequence of this premise, we have a unitary theory with the probability conserved in time, once the inner product in Hilbert's space will have a definite positive norm.

On the other hand, in recent years, it has been observed that the requirements for real energy eigenvalues and the unitarity of the  theory can be obtained in non-Hermitian Hamiltonian operators. Indeed,  in 1998, Bender and Boettcher \cite{Bender} demonstrated that the hermiticity condition can be replaced by a condition analogous to the space-time reflection symmetry, i. e., the $\mathcal{PT}$-symmetry \cite{Bender}. Therefore, if the Hamiltonian operator is invariant under the $\mathcal{PT}$-symmetry operator it will satisfy all the conditions of quantum mechanics mentioned above \cite{Bender, Weigert, Bender1}. If the $\mathcal{PT}$ symmetry is violated in the model, the energy eigenvalues would not be physically acceptable. 

After the publication of the paper of Ref. \cite{Bender}, there has been a significant increase in the interest of many researchers and supporters of the concept of non-Hermitian Hamiltonian. As a matter of fact, this paper allowed a new branch for studies involving these systems, for example, studies in quantum optics \cite{Bender2, Longhi, Razzari}, superconductivity \cite{Rubinstein, Chtc}, lasers \cite{Chong, Liertzer}, microwave cavities \cite{Dietz, Bittner}, electronic circuits \cite{Schindler}, graphene \cite{Fagotti}, among others. However, after the seminal work of Bender \cite{Bender}, interesting developments arise towards the structures of theories with non-Hermitian and non-$\mathcal{PT}$ symmetric Hamiltonians, but with real eigenvalues \cite{swanson, dutra, Ramos}.

In 1948, Claude E. Shannon constructed the concept that would be known as Shannon's entropy or Shannon's Information \cite{Shannon}. Initially, this concept was formulated within the framework of the mathematical theory of communication or information theory \cite{Shannon, Dong, Dong1}. Information theory has allowed the study of cryptography \cite{Grosshans}, coding \cite{Schumacher}, noise theory \cite{Wyner}, and others. In the context of quantum mechanics the interpretation of Shannon's entropy in the space of positions is related to the uncertainty of the location of the particle in space \cite{Dong, Dong1, born}. Similarly, entropy in the momentum space is related to the uncertainty of the particle momentum measurements. Thus, Shannon's entropy presents itself in quantum mechanics as a new formalism for the study of uncertainties and information related to quantum systems \cite{Shannon, Nicolas}.

Theoretical studies on information measures for different quantum systems have been carried out in several studies \cite{Solaimani,Shi11,Yahya,Coffey,Song}. There are many models in which we have the investigation of information theory. For instance, study of systems with position-dependent mass \cite{Serrano,Hua,Yanez11},  and Klein-Gordon oscillator \cite{Boumali}. Information theory is also used to describe models involving the Schr\"odinger equation for several potentials, such as Dirac-delta potential \cite{Bouvrie}, hyperbolic potential \cite{Valencia}, harmonic oscillator in D dimensions and hydrogen atom \cite{Yanez, Dehesa}, Morse and P\"oschl-Teller potentials \cite{Dehesa1, Sun}, infinite potential well \cite{Majernik}, double well \cite{Tserkis},  hydrogen atom confined and free \cite{Murherjee},  Eckart potential \cite{Kumar}, screened Kratzer potential \cite{PPOmadi}, generalized hyperbolic potential \cite{IkotR}, Mobius  square potential \cite{IkotRRR}, exponential-type potential \cite{IkotRRRR}, among others.

The main objective of this work is to present, for the first time in the literature, a study of the influence of an external field on the thermodynamic properties of a non-Hermitian spinless particles ensemble. To achieve our goals, we use the energy spectrum of the system and with the help of Bose-Einstein statistic we build the partition function in order to obtain the thermodynamic properties of the model, i. e., Helmholtz free energy, entropy, free energy and specific heat. We also present, in an unprecedented way, studies about Shannon's entropy on the particles that make up the non-Hermitian ensemble. It is important to remember that for the study of Shannon's entropy we will use wave functions of the model. Clearly, when studying the thermodynamic properties, we consider the ensemble of these spinless particles. In the regime that we study Shannon's entropy we consider only a single particle for investigate the position and moment uncertainties of the model subject to the influence of the external magnetic field. 

This work is organized as follows. In Section II, we present a brief review of the model considered. In section III, we turn our attention to the study of Shannon's entropy and how this quantity changes with the magnetic field. In this way, we present the numerical and graphical results related to Shannon entropy and entropic densities of the model. Finally, in section IV we build a canonical ensemble of these particles and study the thermodynamic properties of the system. Then, we conclude by presenting some results and discussions.

\section{A brief review}

In this study, we will use the model recently introduced by Ramos {\textit et. al.}  \cite{Ramos} and in this section we follow closely that work. We consider a spinless charged particle interacting with the magnetic field. The respective Schr\"odinger equation is written as 
\begin{equation}
i\hbar\partial_{t}\Phi(\vec{r}, t)=\frac{1}{2m}[\hat{p}-e\vec{A}(\vec{r})]^{2}\Phi(\vec{r},t)+V(\vec{r})\Phi(\vec{r},t),
\end{equation}
where $e$ is the electric charge, $m$ is the mass of the particle, $\vec{A}(\vec{r})$ is the electromagnetic vector potential and $V(\vec{r})$ is the scalar potential of the system.

For convenience, we assume that 
\begin{equation}
\label{gauge}
    \vec{A}(\vec{r})=Bx\hat{y},
\end{equation}
where  $B$ is a constant that represent the magnitude of the magnetic field. With the expression (\ref{gauge}), we observe that
\begin{equation}
    \vec{B}(\vec{r})=\vec{\nabla}\times\vec{A}(\vec{r})\rightarrow \vec{B}(\vec{r})=B\hat{z}.
\end{equation}

Therefore, the electron moves in a plane perpendicular to the magnetic field, which gives rise to a spectrum of energy well known in the specialized literature as the levels of Landau \cite{Griffiths, Ramos}.

We know that when we work with non-relativistic quantum systems, it is necessary to formulate a Hermitian Hamiltonian operator. However, as discussed by Ramos {\it et. al} \cite{Ramos}, it is possible to describe some systems with non-Hermitian operators that have real eigenvalues. In other words, although the operator is non-Hermitian, these operators can describe physical aspects of the systems. 

As our start point, we will assume a potential of the form \cite{Ramos}:
\begin{equation}
\label{potential}
    V(\vec{r})=i\sqrt{2}\alpha\hat{r}\cdot[\hat{p}-e\vec{A}],
\end{equation}
with $\hat{r}=\frac{1}{\sqrt{2}}(\hat{x},\hat{y},0)$ and $\alpha$ a real parameter.

Observing Eq. (\ref{potential}), we notice that the potential is not invariant under $\mathcal{PT}$  transformations. Thus, we note that the Hamiltonian operator that describes the Landau system is non-Hermitian and is not symmetric by $ \mathcal{PT}$.

Assuming variable separation $\Phi(\vec{r},t)=$e$^{-i\mathcal{E}t/\hbar}\psi(x,y)$ in the model, we observe that
\begin{align}
-\frac{\hbar^{2}}{2m}[\partial_{x}^{2}\psi+\partial_{y}^{2}\psi]&+i\frac{\hbar e B}{m}x\partial_{y}\psi
+\frac{e^{2}B^{2}}{2m}x^{2}\psi+\alpha\hbar[\partial_{x}\psi+\partial_{y}\psi]-i\alpha eBx\psi=\mathcal{E}\psi.
\end{align}

The equation above is independent of the $y$ variable. From the fundamental of quantum mechanics, this implies that the momentum operator component $\hat{p}_{y}$ commutes with the Hamiltonian. In other words, this physical quantity is a conserved one. Indeed, it worth to comment here that the system is effectively bidimensional. Then, the solution of the system can be written as  $\psi=$e$^{i\frac{p_{y}}{\hbar}y}f(x)$ with $-\infty <p_{y}<\infty$ corresponding to the eigenvalue associated with the momentum operator in the $y$-component.

Rewriting the previous equation, now we have for the function $f(x)$,
\begin{align}
\label{wavex}
    f''-\frac{2m\alpha}{\hbar}f'-\frac{m^{2}\omega^{2}}{\hbar^{2}}&\bigg(x-\frac{p_{y}}{m\omega}\bigg)^{2}f+i\frac{2m\alpha eB}{\hbar^{2}}\bigg(x-\frac{p_{y}}{m\omega}\bigg)f+\frac{2m\mathcal{E}}{\hbar^{2}}f=0,
\end{align}
where $\omega=eB/m$ is the frequency of the cyclotron. We assume the parameters
\begin{equation}
\beta=\frac{2m\mathcal{E}}{\hbar^{2}} \hspace{1cm} \text{and} \hspace{1cm} \theta=\frac{2m\alpha}{\hbar}.
\end{equation}

In addition, we consider the respective changes in the independent and dependent variables as
\begin{equation}
\label{transformation}
    \xi=\sqrt{\frac{m\omega}{\hbar}}\bigg(x-\frac{p_{y}}{m\omega}-i\frac{\theta\hbar}{2m\omega}\bigg),
\end{equation}
and
\begin{equation}
\label{transformation1}
    f(\xi)=\xi\text{e}^{-\frac{\xi^{2}}{2}}\text{e}^{\frac{\theta}{2}\sqrt{\frac{\hbar}{m\omega}}\xi}F(\xi).
\end{equation}

Replacing the equations (\ref{transformation}) and (\ref{transformation1}), we reduced the expression (\ref{wavex}) to
\begin{equation}
\label{wavex1}
    F''(\xi)+\bigg(\frac{2}{\xi}-2\xi\bigg)F'(\xi)+\bigg[\frac{\beta\hbar}{4m\omega}-\frac{\theta^{2}\hbar}{8m\omega}-\frac{3}{4}\bigg]F(\xi)=0.
\end{equation}

Now after we make $\zeta=\xi^{2}$, we obtain the equation known as confluent hypergeometric equation, namely,
\begin{equation}
    \zeta F''(\zeta)+\bigg(\frac{3}{2}-\zeta\bigg)F'(\zeta)+\bigg[\frac{\beta\hbar}{4m\omega}-\frac{\theta^{2}\hbar}{8m\omega}-\frac{3}{4}\bigg]F(\zeta)=0.
\end{equation}
 
Remembering that $\Phi(\vec{r}\rightarrow 0,t)\rightarrow 0$, we finally obtain that the wave function is described by
\begin{align}
\label{solution}\nonumber
    \Phi_{n}(\vec{r},t)=& A\sqrt{\frac{m\omega}{\hbar}}\bigg(x-\frac{p_{y}}{m\omega}-i\frac{\theta\hbar}{2m\omega}\bigg)\text{e}^{-i\frac{\mathcal{E}}{\hbar}t}\text{e}^{i\frac{p_{y}}{\hbar}y}\text{e}^{-\frac{m\omega}{2\hbar}\bigg(x-\frac{p_{y}}{m\omega}-i\frac{\theta\hbar}{2m\omega}\bigg)^2}\text{e}^{\frac{\theta}{2}\bigg(x-\frac{p_{y}}{m\omega}-i\frac{\theta\hbar}{2m\omega}\bigg)} \times \\ &_{1}F_{1}\bigg[-n,\frac{3}{2};\frac{m\omega}{\hbar}\bigg(x-\frac{p_{y}}{m\omega}-i\frac{\theta\hbar}{2m\omega}\bigg)^2\bigg], 
\end{align}
{\bf with $n=0,1,2,...$ and $A$ is a constant of normalization of the wave 
function.}

The normalization constant may be found for each energy level by the expression:
\begin{align}
\label{normalization}
    \int_{-\infty}^{\infty}|\Phi_{n}(\vec{r},t)|^{2}=1.
\end{align}

Considering the ground state, i. e., $n=0$ in eq. (\ref{solution}) and using the expression (\ref{normalization}), we have that $\Phi_{0}(\vec{r}, t)$ is
\begin{align}
     \Phi_{0}(\vec{r},t)=& A_{0}\bigg(x-\frac{p_{y}}{m\omega}-i\frac{\theta\hbar}{2m\omega}\bigg)\text{e}^{-i\frac{\mathcal{E}}{\hbar}t}\text{e}^{i\frac{p_{y}}{\hbar}y}\text{e}^{-\frac{m\omega}{2\hbar}\bigg(x-\frac{p_{y}}{m\omega}-i\frac{\theta\hbar}{2m\omega}\bigg)^2}\text{e}^{\frac{\theta}{2}\bigg(x-\frac{p_{y}}{m\omega}-i\frac{\theta\hbar}{2m\omega}\bigg)}, 
\end{align}
where,
\begin{align}\nonumber
A_{0}=&\frac{16 m^2 \omega^2}{\sqrt{\pi } \left[\hbar  \left(5 \theta ^2 \hbar +8 m \omega \right)+16 (m-1)^2 p_{y}^2+8 \theta  (m-1) p_{y} \hbar \right]}\text{e}^{-\frac{\theta ^2 \hbar ^2+16 (m-1) m p_{y}^2+4 \theta  \hbar  [\theta  m+2 (m-1) p_{y}]}{16 m \omega  \hbar }}.
\end{align}

Due to the complexity of the wave function, analytical demonstration of other normalized states is not viable. As the objective of our work consists of studying the thermodynamic properties and Shannon's entropy, we will numerically normalize the other energy states.

\begin{figure}[h!]
\centering
\includegraphics[scale=0.4]{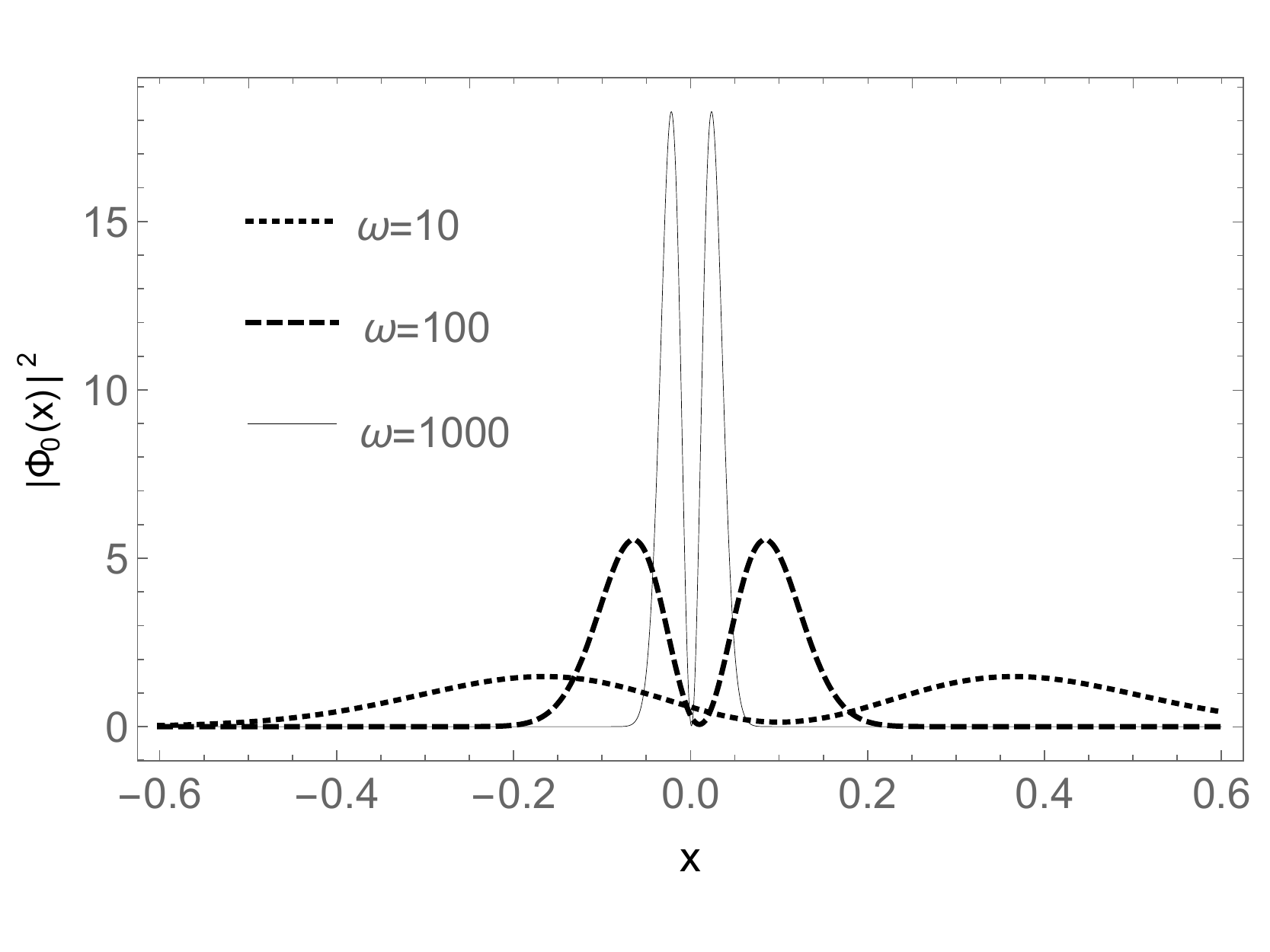}
\includegraphics[scale=0.4]{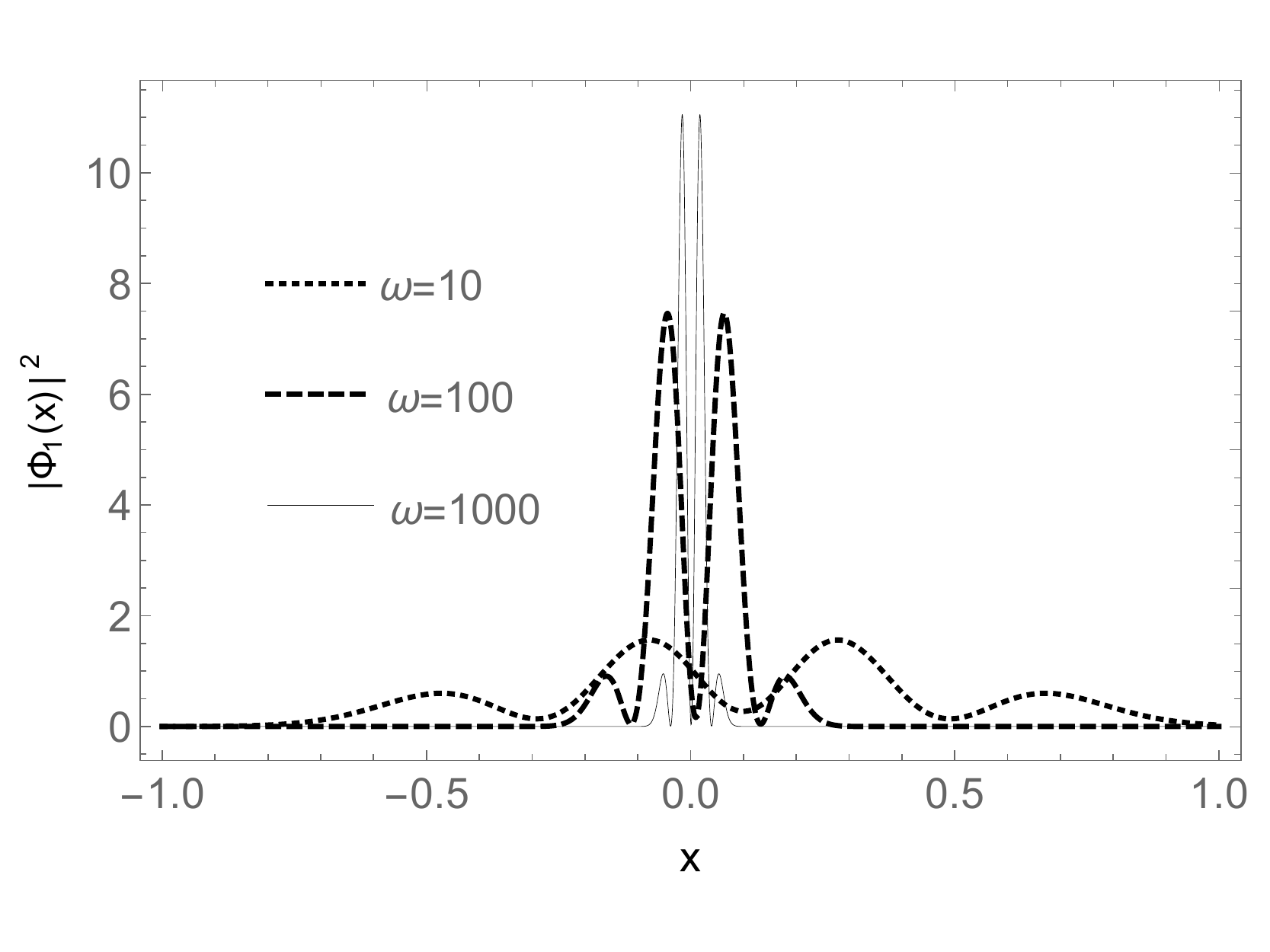}
\caption{Behavior of probability densities in the position space for the ground state $(n=0)$ and for the first excited state $(n=1)$.} \label{fig00}
\end{figure}

Numerically analyzing the behavior of the wave function normalized for the ground state and for the first excited state, we will have the graphical results presented in the figure (\ref{fig00}).

Corresponding eigenvalues of energy are given by
\begin{align}
\label{energyM}
\mathcal{E}_{n}=\bigg(2n+\frac{3}{2}\bigg)\hbar\omega+m\alpha^{2}.
\end{align}
From the expression above, we notice that the energy spectrum has a linear dependence with the magnetic field for any value of $n$. If the magnetic field is null, the model will have infinite degenerate solutions with eigenvalue $ \mathcal{E}_{n}=m\alpha^2$.

We observed that this model is similar to a harmonic oscillator. In this way, the thermodynamic properties are essentially the properties of the harmonic oscillator, as discussed in ref. \cite{DongR}.


\section{Information theory: Shannon's entropy}

Based on thermodynamics, the Shannon entropy concept provides the irreversibility of the physical system. According to statistical physics, the measure of entropy appears as a measure associated with the degree of disorder of the system.

Shannon's concept of entropy was first proposed in 1948 by Claude E. Shannon in the paper entitled  \textit{A mathematical theory of communication} \cite{Shannon}. In this work, Shannon describes entropy as an element of information theory. In this context, entropy is a measure of uncertainty in a given probability distribution, so it is called Shannon's entropy or information entropy.

The entropy can be interpreted as a starting point for measuring the uncertainty of a probability distribution associated with an information source. According to Born \cite{born}, the statistical interpretation of a quantum system is described by the probability density that is defined by
\begin{equation}
    \rho(x)\equiv |\Phi(x,t)|^{2}.
\end{equation}

Thus, for a probability density of a continuous system in one-dimensional space, Shannon's entropy for the n-th state is described by
\begin{equation}
\label{shannonx}
    S_{x}^{n}=-\int_{-\infty}^{\infty}|\Phi_{n}(x,t)|^{2}\text{ln}(|\Phi_{n}(x,t)|^{2})\, dx,
\end{equation}
while in the momentum-k space it is
\begin{equation}
    S_{k}^{n}=-\int_{-\infty}^{\infty}|\Phi_{n}(k,t)|^{2}\text{ln}(|\Phi_{n}(k,t)|^{2})\, dk,
\end{equation}
where $\Phi(k,t)$ is the wave function in the reciprocal space or in the momentum space \cite{Griffiths}.

\subsection{Numerical result of the information}

From now on, we analyzed Shannon's information of a charged non-Hermitian spinless particle interacting with the magnetic field. In this context, we investigate how information is modified due to the influence of the magnetic field on the model. For this, we analyzed the wave functions for the first energy levels of the model, that is, $n=0,1,2,3$. With the wave function described for the first energy levels of the model, we can observe the influence of the magnetic field on the uncertainties related to the solutions. With this in mind, we replace the normalized wave eigenfunctions described by the expression (\ref{solution}) of the first energy levels in the expression (\ref{shannonx}), and later through the Fourier transform
\begin{equation}
    \Phi_{n}(k,t)=\frac{1}{\sqrt{2\pi}}\int\,\Phi_{n}(x,t)\,\text{e}^{ikx}\,dx.
\end{equation}
We find that the wave eigenfunction in reciprocal space for the first energy levels have the following form:
\begin{align}
    \Phi_{0}(k)=A_{1}\frac{i(2k+i\theta)\hbar}{m\omega\sqrt{\pi}}\text{e}^{\frac{8ikm\omega p_y-4km\omega\theta\hbar-4k^2 m\omega\hbar-4ikm\theta\omega\hbar+m\theta^{2}\omega\hbar}{8m^2\omega^{2}}};
\end{align}

\begin{align}
    \Phi_{1}(k)=A_{2}\frac{(-2ik+\theta)\hbar[6m\omega-(2ki\theta)^{2}\hbar]}{6m\omega^{2}\sqrt{\pi}}\text{e}^{\frac{8ikm\omega p_y-4km\omega\theta\hbar-4k^2 m\omega\hbar-4ikm\theta\omega\hbar+m\theta^{2}\omega\hbar}{8m^2\omega^{2}}};
\end{align}

and
\begin{align}
    \phi_{3}(k)=A_{3}\frac{i(2k+i\theta)\hbar[60m\omega^{2}-20m\omega(2k+i\theta)^{2}\hbar+(2k+i\theta)^{4}\hbar^{2}]}{60m\omega^{3}\sqrt{\pi}}\text{e}^{\frac{8ikm\omega p_y-4km\omega\theta\hbar-4k^2 m\omega\hbar-4ikm\theta\omega\hbar+m\theta^{2}\omega\hbar}{8m^2\omega^{2}}},
\end{align}
where $A_{1}$, $A_{2}$ and $A_{3}$, are normalization parameter.

\begin{figure}[h!]
\centering
\includegraphics[scale=0.4]{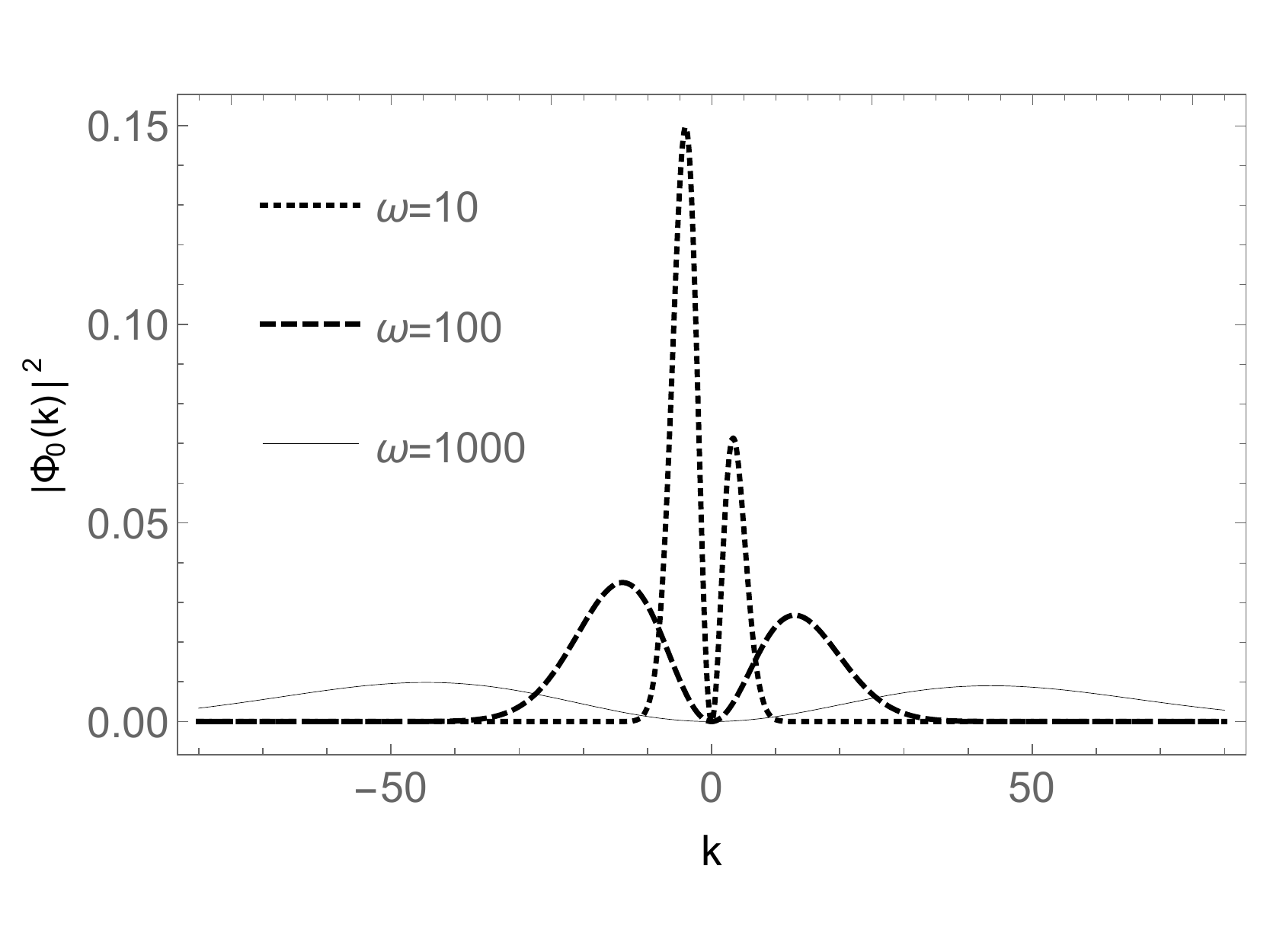}
\includegraphics[scale=0.4]{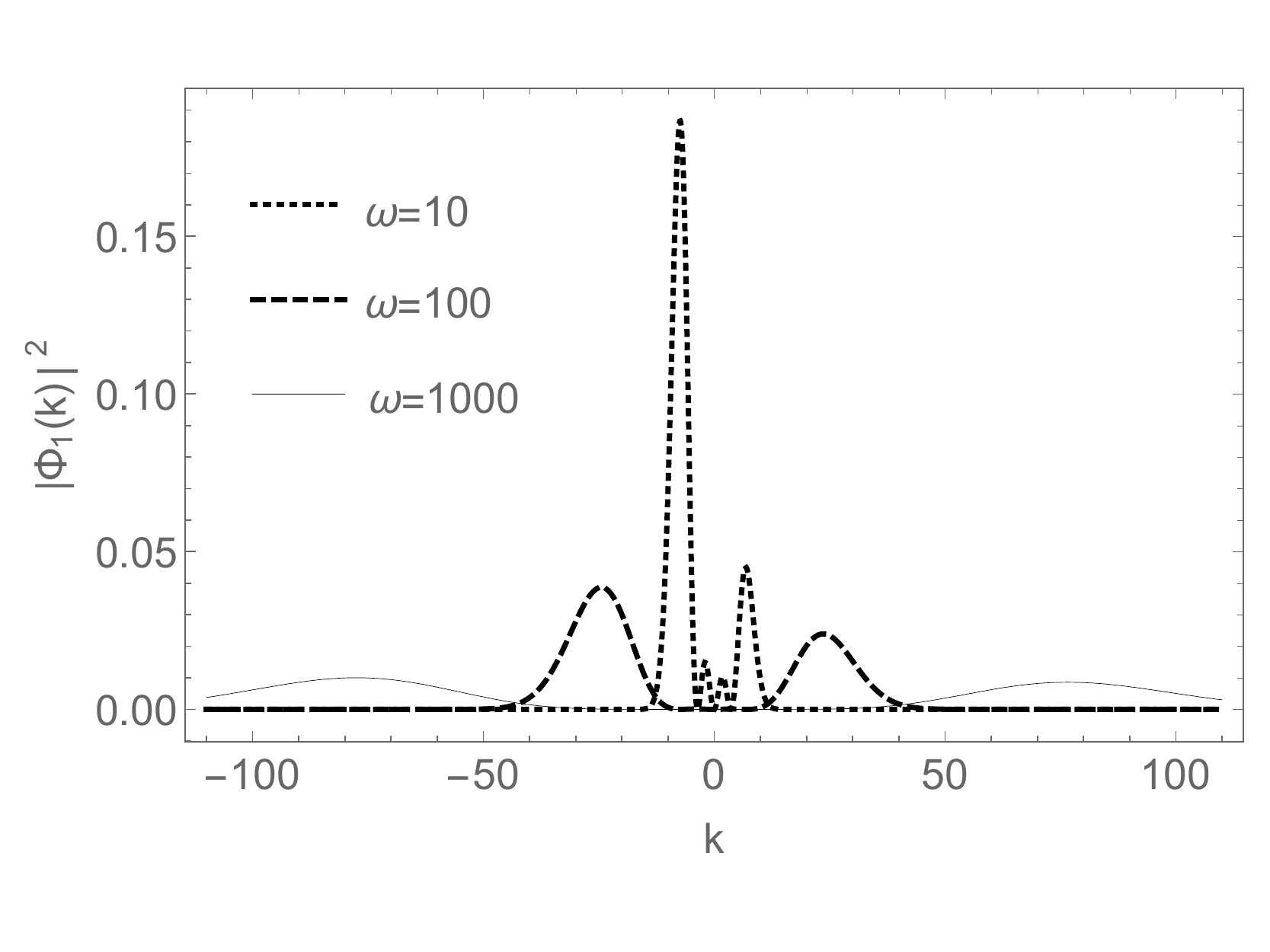}
\caption{Behavior of probability densities in the reciprocal space for the ground state $(n=0)$ and for the first excited state $(n=1)$.} \label{waveM}
\end{figure}

We emphasize that the expressions for the other states are more complicated to work. Therefore, we will work with numerical eigenfunction for carry out the study of Shannon's entropy of the model. In figure (\ref{waveM}) we show the behavior of the first order normalized wave functions in the reciprocal space.

On the other hand, using wave eigenfunctions in the coordinate space and reciprocal space, we study the Shannon entropy of the model and investigate the influence of the external field in this quantity.

Let us now to recall the work of  Bialynicki-Birula and Mycielski (BBM) \cite{BBM}. In that work they derive a new uncertainty relation in quantum mechanics. As a matter of fact, this new relation has an interpretation in terms of information theory. Thus, the entropic uncertainty relationship that relates the position and momentum uncertainties must respect the relation 
\begin{equation}
 S_{x}^{n}+S_{k}^{n}\geq d[1+\text{ln}(\pi)],
 \end{equation}
 where $d$ denotes the dimension of the position and momentum space  \cite{BBM}.
 
Although the system is effectively spatially unidimensional, since the variable $y$ is cyclic, the solution in this coordinate will not contribute to the measurement of information or uncertainties of the physical system. Therefore the equation above turns to
\begin{equation}
    S_{x}^{n}+S_{k}^{n}\geq [1+\text{ln}(\pi)]\cong 2.14473.
\end{equation}

Numerically calculating the quantities $S_{x}^{n}$ and $S_{k}^{n}$, we obtain the results as the cyclotron frequency increases (Table I). In other words, we obtain the information when magnetic field becomes stronger.

\begin{table}[h!]
\centering
\begin{tabular}{|c|c|c|c|c|c|}\hline
\hline
$n$ & $\omega$ & $S_{x}$ & $S_{k}$ & $S_{x}+S_{k}$ \\ \hline
\hline
  & 10   & 0.07511  & 2.58794  & 2.66305 \\
0 & 100  & -1.24617 & 3.93031  & 2.68415 \\
  & 1000 & -2.44074 & 5.12615  & 2.68541 \\ \hline
  & 10   & 0.29949  & 2.51618  & 2.81568 \\
1 & 100  & -1.33164 & 3.91697  & 2.58533 \\
  & 1000 & -2.59974 & 5.15778  & 2.55804 \\ \hline
  & 10   & 0.34254  & 2.33426  & 2.67681 \\
2 & 100  & -1.35202 & 3.89831  & 2.5463  \\
  & 1000 & -2.62173 & 5.16562  & 2.54389  \\ \hline
  & 10   & 0.35207  & 2.19412  & 2.54619  \\
3 & 100  & -1.35095 & 3.88301  & 2.53205  \\
  & 1000 & -2.62896 & 5.16816  & 2.53919  \\ \hline
\end{tabular}\\
\caption{Numerical result of the Shannon's entropy in the system for several values of the cyclotron frequency.}
\end{table}

\begin{figure}[h!]
\centering
\includegraphics[scale=0.7]{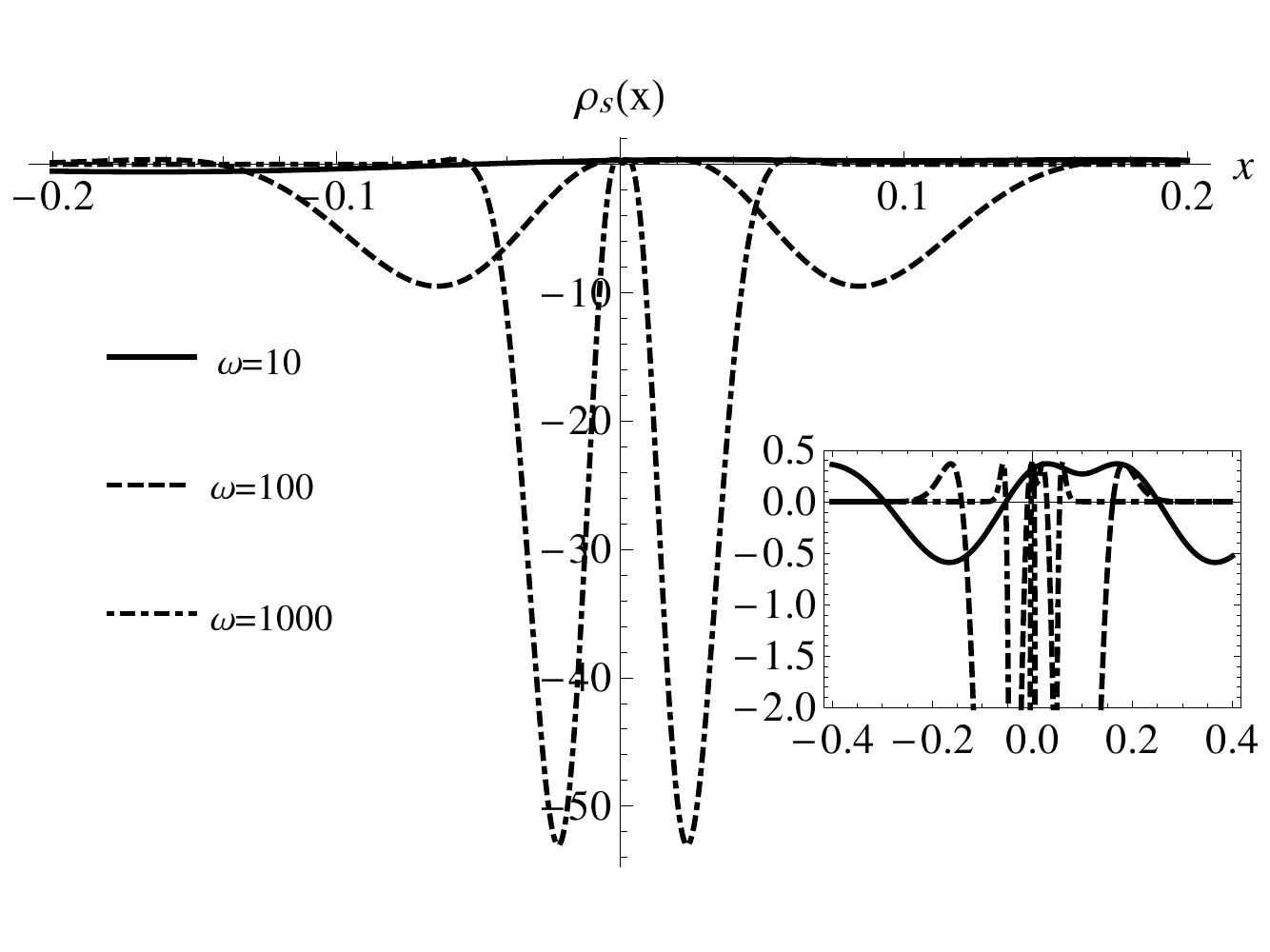}
\caption{Entropic density in the position space when $n=0$ for several
values of the cyclotron frequency.} \label{fig1}
\end{figure}

\begin{figure}[h!]
\centering
\includegraphics[scale=0.7]{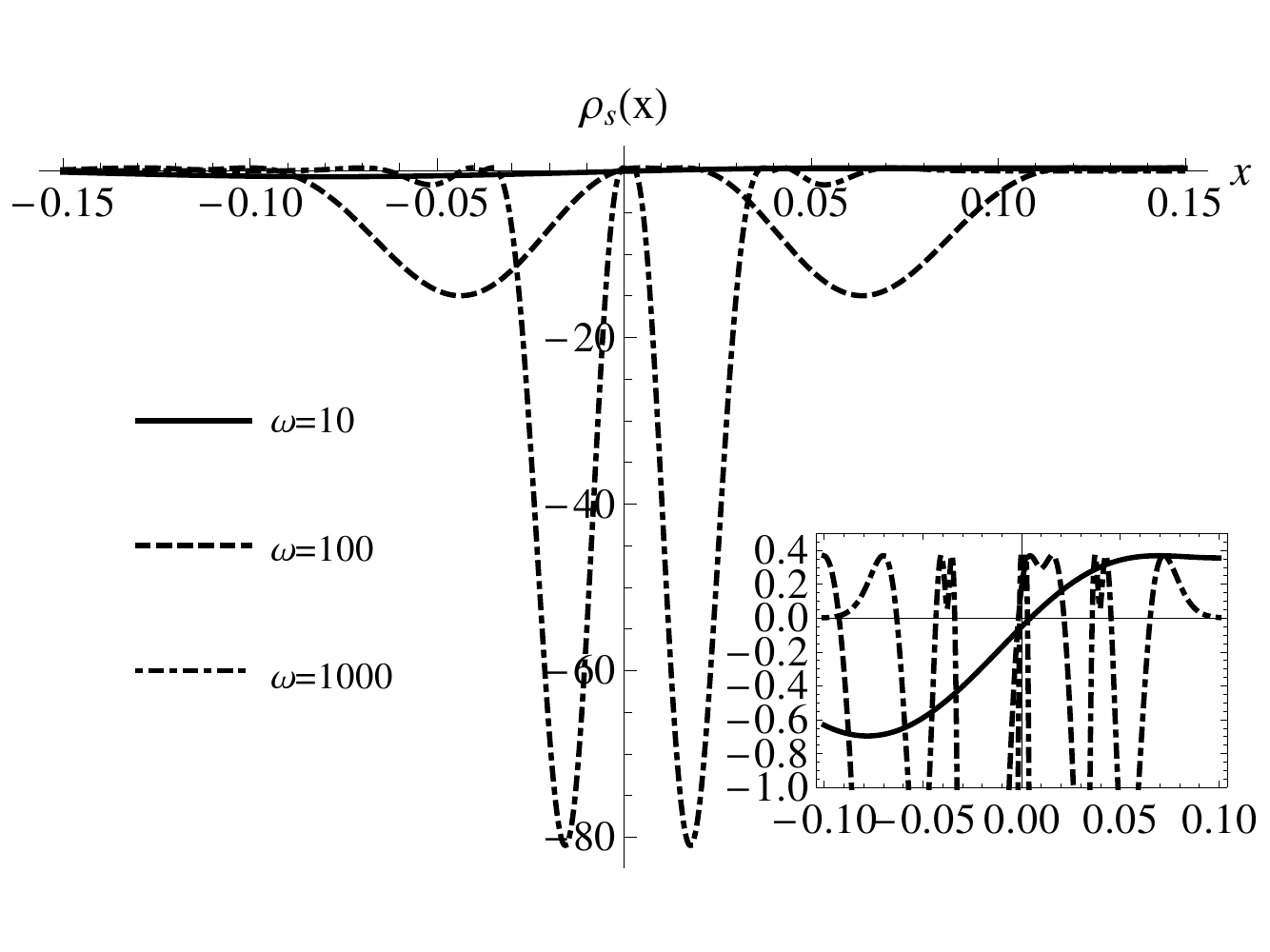}
\vspace{-10pt}
\caption{Entropic density in the position space when $n=1$ for several
values of the cyclotron frequency.} \label{fig2}
\end{figure}

\begin{figure}[h!]
\centering
\includegraphics[scale=0.7]{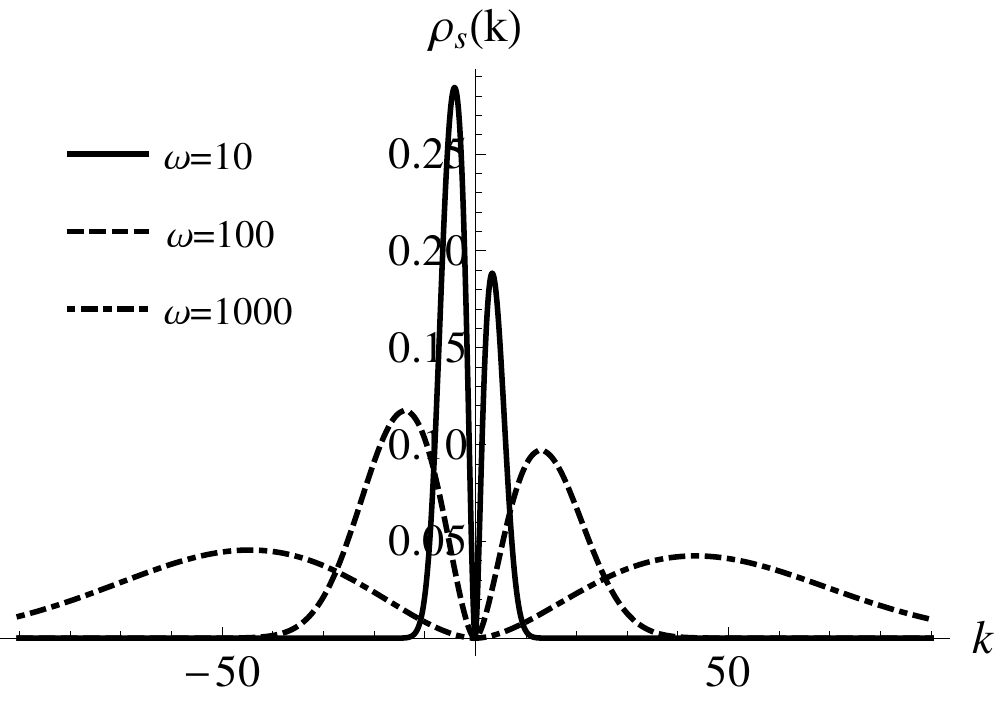}
\vspace{-10pt}
\caption{Entropic density in the momentum space when $n=0$ for several
values of the cyclotron frequency.} \label{fig3}
\end{figure}

\begin{figure}[h!]
\centering
\includegraphics[scale=0.7]{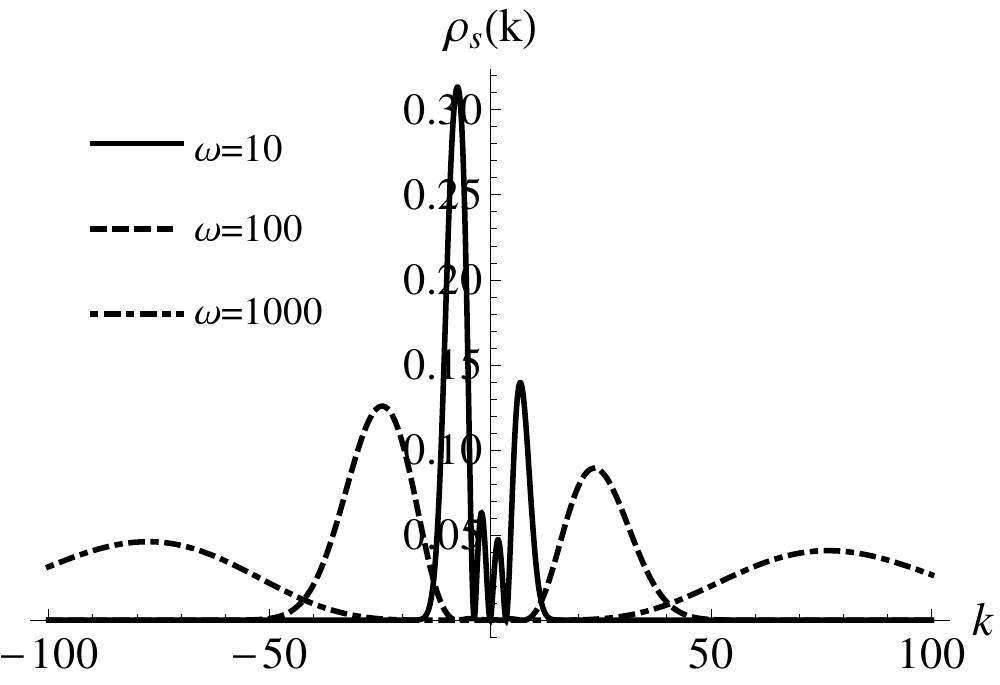}
\vspace{-10pt}
\caption{Entropic density in the momentum space when $n=1$ for several values of the cyclotron frequency.} \label{fig4}
\end{figure}

The Shannon's information can be well illustrated through the graphics of the entropic density in the position space and in the momentum space for several values of the magnetic field (cyclotron frequency). The results are shown in Figs.
 (\ref{fig1}), (\ref{fig2}), (\ref{fig3}) and (\ref{fig4}).

\vspace{15pt}

\begin{figure}[h!]
\centering
\includegraphics[scale=0.6]{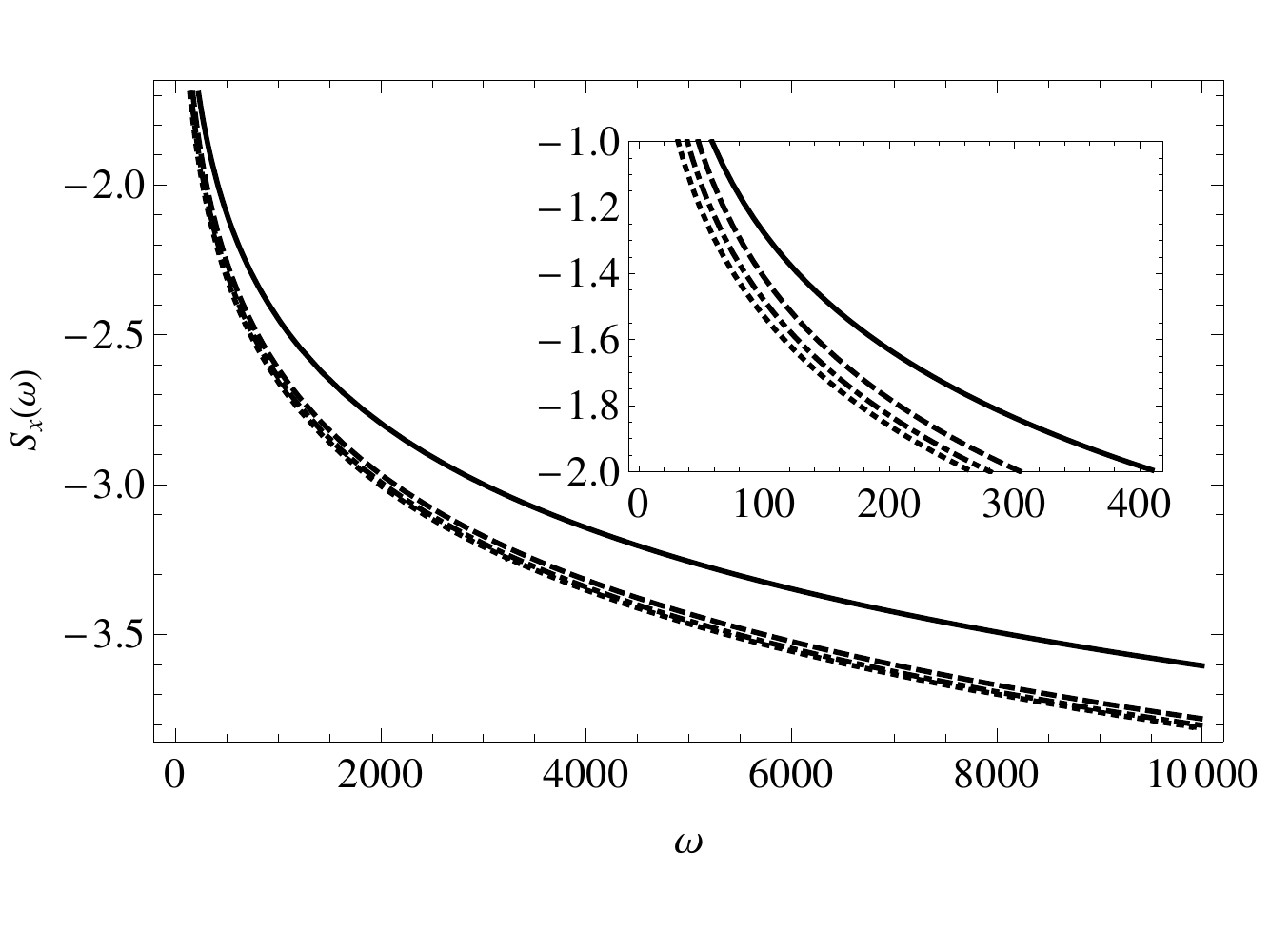}
\vspace{-10pt}
\caption{Shannon's entropy in the position space in function of the cyclotron frequency. The energy states are described for $n=0$ (continuous line), $n=1$ (dashed line), $n=2$ (dotted-dashed) and $n=3$ (dotted line).} \label{fig5}
\end{figure}

\begin{figure}[h!]
\centering
\includegraphics[scale=0.6]{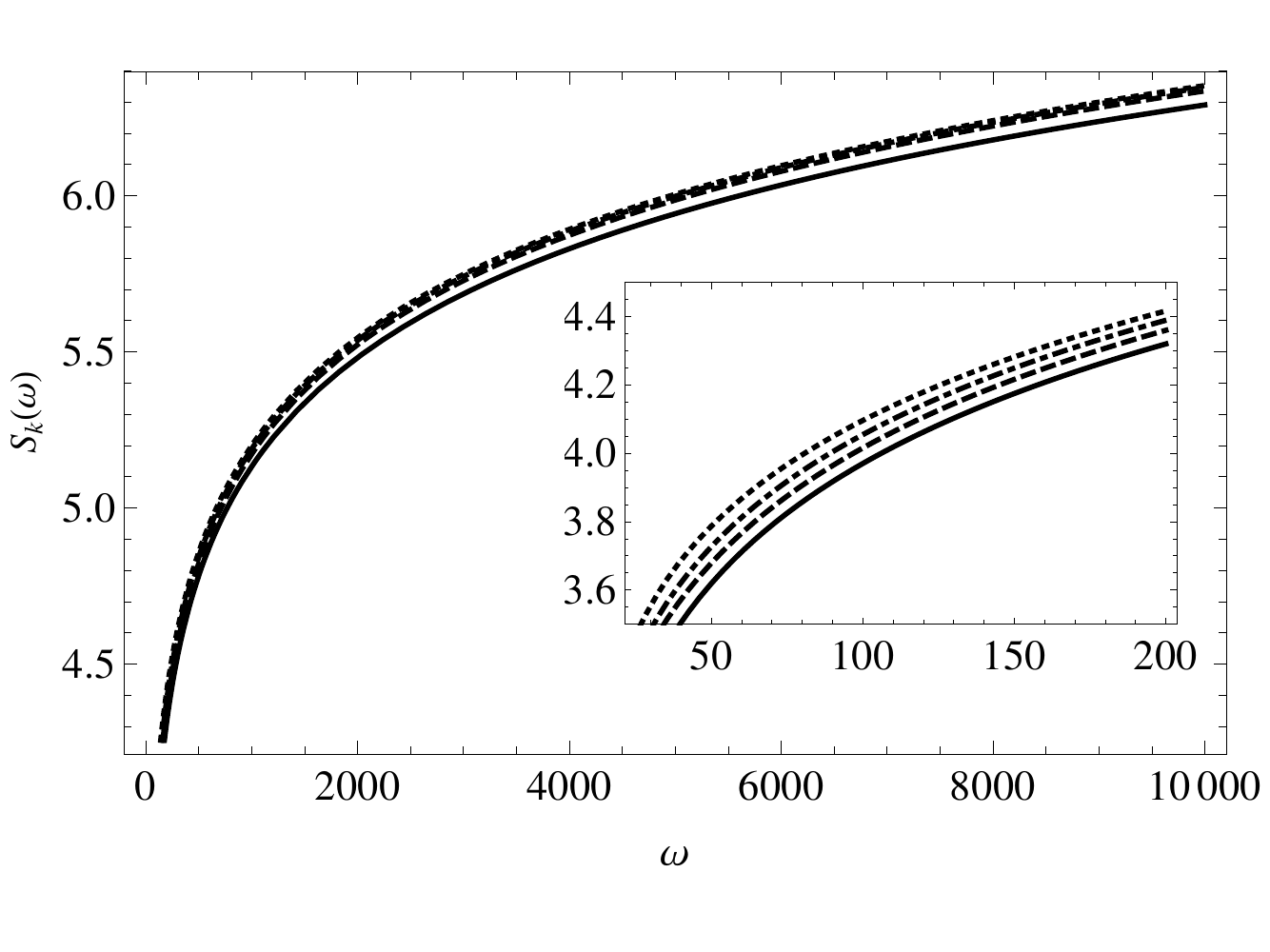}
\vspace{-10pt}
\caption{Shannon's entropy in the momentum space in function of the cyclotron frequency. The energy states are described for $n=0$ (continuous line), $n=1$ (dashed line), $n=2$ (dotted-dashed) and $n=3$ (dotted line).} \label{fig6}
\end{figure}

We can make an analysis of the numerical results for Shannon's entropy in the table (I). We interpolate the functions that represent Shannon's information and present the graphical results as a function of the cyclotron frequency. The results are shown in Figs. (\ref{fig5}) and (\ref{fig6}).

\section{Thermodynamic properties}
Before starting our study, it is interesting to mention that in the current literature several interesting work discussing the thermodynamic properties of quantum systems has emerged in recent years. Among these works, we have studies of thermal properties of diatomic molecules in the presence of external magnetic fields \cite{Rampho}, study of Denf-Fan-Eckart model \cite{Edet1}, study of magnetic susceptibility of the Hellman potential in Aharanov-Bohm flux with finite temperature \cite{Edet}, study thermal of the Poschl-Teller model \cite{Edet2}, study thermal on potential pseudo-harmonic subject models in external fields \cite{Ikot11}, among others.

In this way, motivated by some of these contributions, we will study the thermodynamic properties of the system non-Hermitian particle ensemble. For this, we built a canonical ensemble consisting of spinless particles and under the influence of the magnetic field. Initially, for the calculation of the properties we start by defining the well known partition function \cite{Okorie}. We recall that the partition function plays a fundamental role in statistical mechanics, where all thermodynamic quantities are constructed from it. Since the model is not degenerate, the partition function that describes the accessible states of the particles in a thermal bath \cite{Okorie,Ikot}, is described by
\begin{equation}
    Z_{1}=\sum_{n=0}^{\infty}\text{e}^{-\beta\mathcal{E}_{n}}, 
\end{equation}
where $\beta=1/k_{B}T$; $k_{B}$ is the Boltzmann constant and $T$ is the temperature of the ensemble in thermodynamic equilibrium. For later convenience in the analysis of the graphical results, we will assume the variable $ \tau=k_{B}T$. 

After obtaining the partition function, it is possible to study all the Boltzmann thermodynamic properties of the model. In this work, we will turn our attention to the study of the main thermodynamic functions and the influence of the magnetic field on their properties. Namely, the main properties are the Helmholtz free energy $\mathcal{F}(\tau)$, the mean energy $\mathcal{U}(\tau)$, the entropy $\mathcal{S}(\tau)$ and the heat capacity $\mathcal{C}_{v}(\tau)$. Mathematically, these quantities for a canonical ensemble are given by
\begin{align}
 &\mathcal{F}=-\frac{1}{\beta}\text{ln}\, Z_{N}; \hspace{1cm} \mathcal{U}=-\frac{\partial}{\partial\beta}\text{ln}\, Z_{N}; \hspace{1cm} \mathcal{S}=k_{B}\beta^{2}\frac{\partial\mathcal{F}}{\partial \beta}; \hspace{1cm} \mathcal{C}_{v}=-k_{b}\beta^{2}\frac{\partial\mathcal{U}}{\partial\beta},
\end{align}
where $Z_{N}$ is the total partition function and $N$ is the amount of particles that make up the ensemble.

\subsection{The results}

Now, we consider Eq. (\ref{energyM}) to build the partition function of the model. Then, we obtain that 
\begin{equation}
\label{partition}
    Z_{1}=\text{e}^{-m\alpha^{2}\beta}\sum_{n=0}^{\infty}\text{e}^{-(2n+\frac{3}{2})\hbar\omega\beta}=\frac{\text{e}^{-(m\alpha^{2}+\frac{3}{2}\hbar\omega)\beta}}{1-\text{e}^{-2\hbar\omega\beta}}.
\end{equation}

From the partition function, we obtain the main thermodynamic functions. Namely, 
\begin{align}
    \mathcal{F}=-\frac{N}{\beta}\text{ln}\bigg(\frac{\text{e}^{-(m\alpha^{2}+\frac{3}{2}\hbar\omega)\beta}}{1-\text{e}^{-2\hbar\omega\beta}}\bigg),
\end{align}

\begin{align}
    \mathcal{U}=\frac{2m\alpha^{2}N(\text{e}^{-2\hbar\omega\beta}-1)-(3+\text{e}^{-2\hbar\omega\beta})\hbar\omega N}{2(\text{e}^{-2\hbar\omega\beta}-1)},
\end{align}

\begin{align}\nonumber
    \mathcal{S}&=Nk_{B}m\alpha^{2}\beta+\frac{(3+\text{e}^{-2\hbar\omega\beta})\hbar\omega\beta k_{B}N}{2(1-\text{e}^{-2\hbar\omega\beta})} +Nk_{B}\text{ln}\bigg(\frac{\text{e}^{-(m\alpha^{2}+\frac{3}{2}\hbar\omega)\beta}}{1-\text{e}^{-2\hbar\omega\beta}}\bigg),
\end{align}
and
\begin{equation}
\mathcal{C}_{v}=\frac{4\text{e}^{-2\hbar\omega\beta}N\hbar^{2}\omega^{2}\beta^{2}k_{B}}{(\text{e}^{-2\hbar\omega\beta}-1)^{2}}.    
\end{equation}

Analyzing the behavior of properties when the frequency $\omega\rightarrow 0 $. We note that, $\mathcal{F}(\omega\rightarrow 0)\rightarrow -\infty$, $\mathcal{U}(\omega\rightarrow 0)\rightarrow 0$,
$\mathcal{S}(\omega\rightarrow 0)\rightarrow\infty$ and $\mathcal{C}_{v}(\omega\rightarrow 0)\rightarrow 0$.

\begin{figure}[h!]
\centering
\includegraphics[scale=0.52]{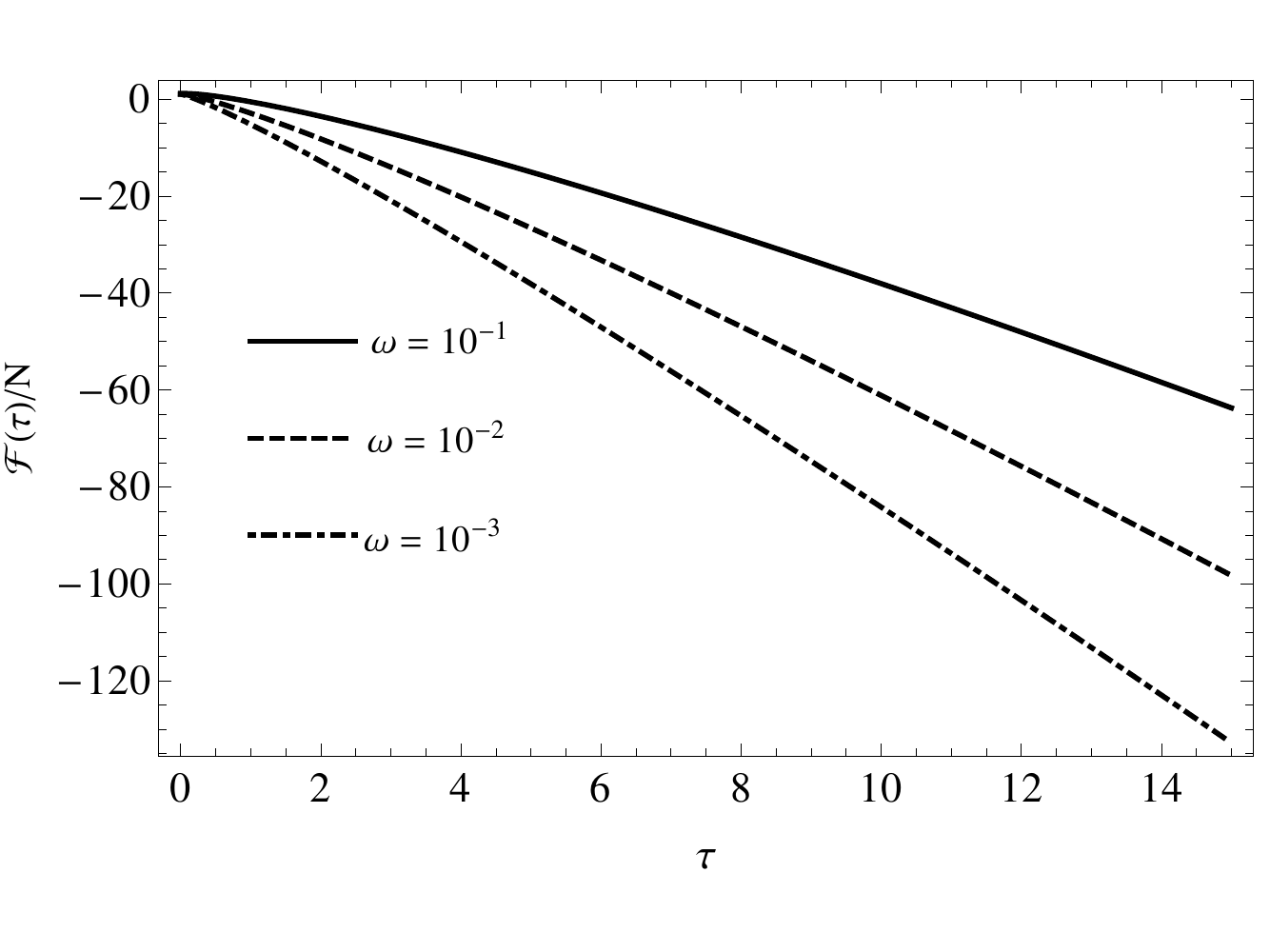}
\includegraphics[scale=0.52]{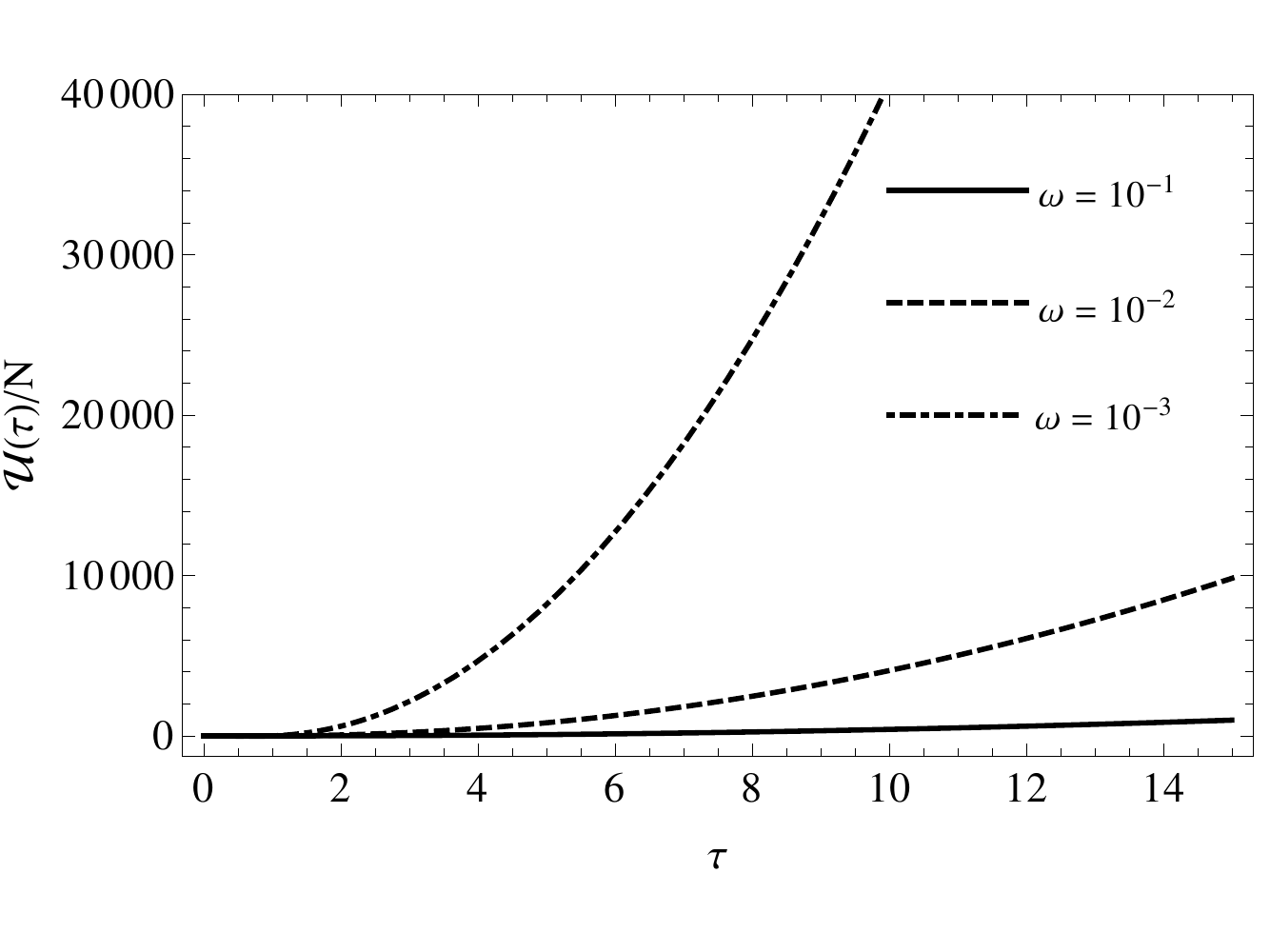}\\
\includegraphics[scale=0.52]{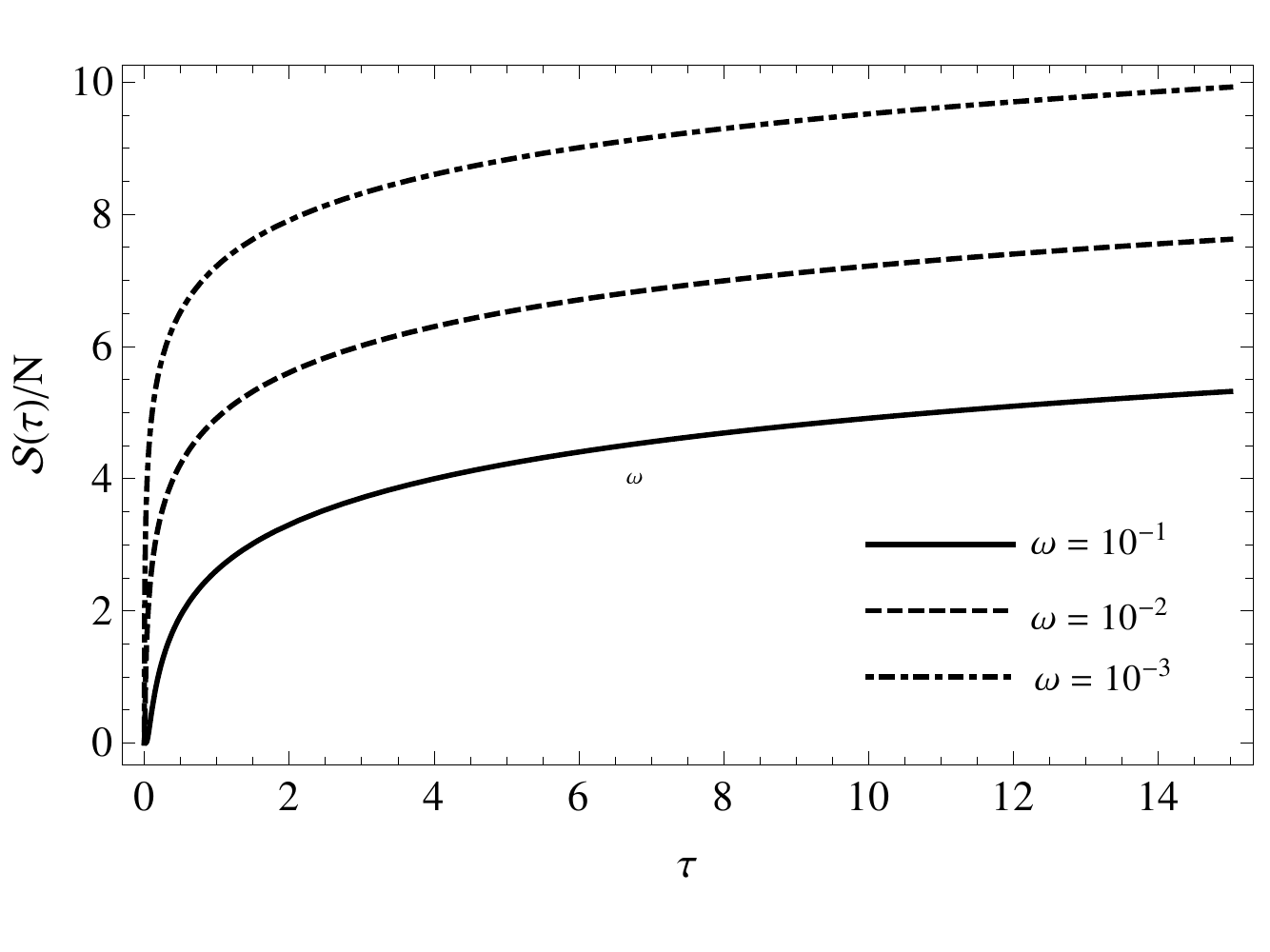}
\includegraphics[scale=0.52]{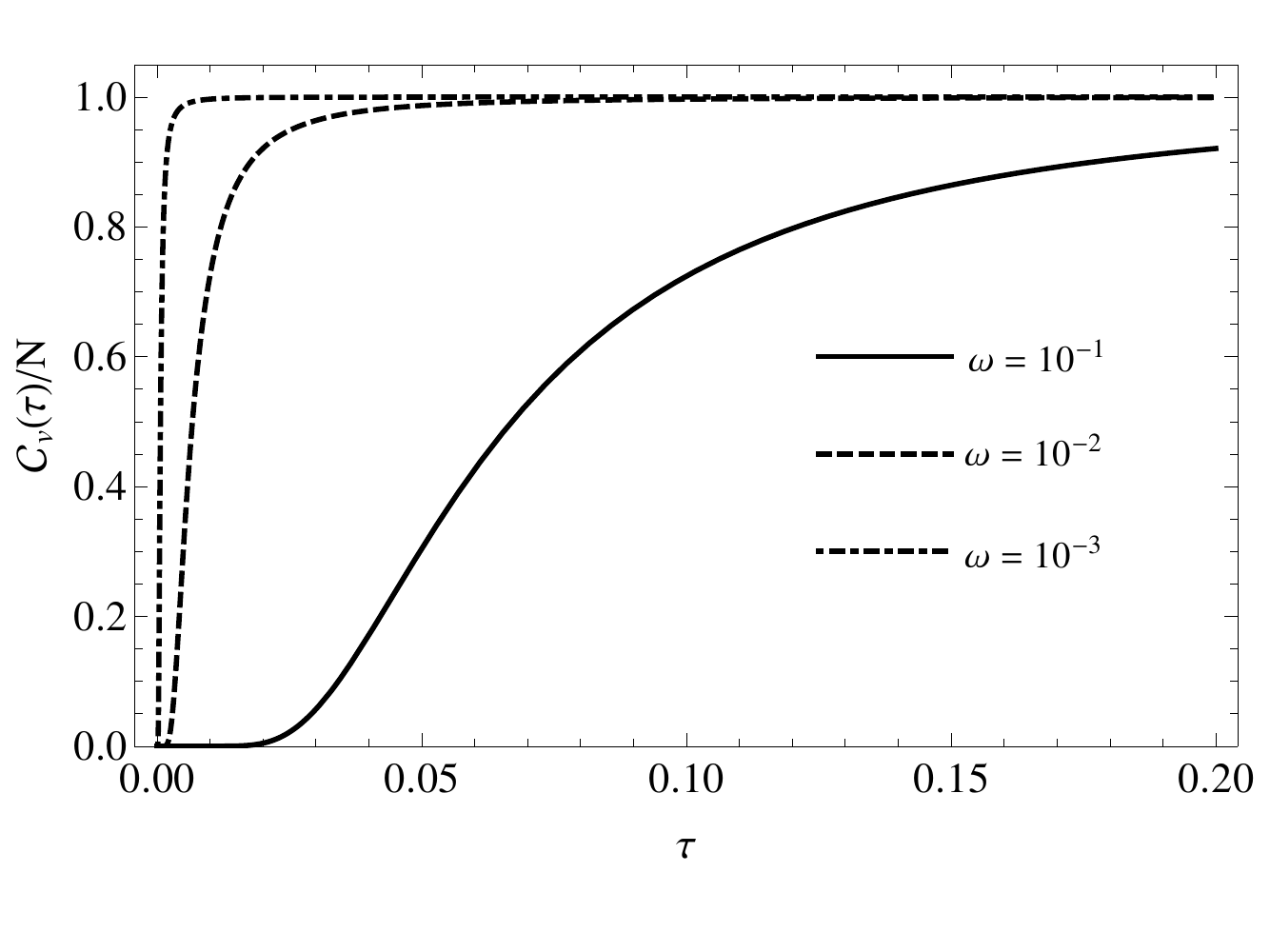}
\vspace{-10pt}
\caption{ Helmholtz free energy $\mathcal{F}(\tau)$, the mean energy $\mathcal{U}(\tau)$, the entropy $\mathcal{S}(\tau)$ and the heat capacity $\mathcal{C}_{v}(\tau)$, for several value of the cyclotron frequency.} \label{fig7}
\end{figure}

The thermodynamic functions are displayed together in Fig. (\ref{fig7}). Subsequently, in the concluding remarks, we present a detailed analysis of these graphs. Also, we discuss the results of the information theory obtained for the system.


\section{Results and discussions}

Through the numerical results presented in table-1 and the graphical results presented in figures $ (3, 4, 5, 6) $,we noticed that the information tends to decrease in the space of the positions with increasing field. In contrast, it increases in the space of momentum, respecting the entropic uncertainty relationship. We also concluded that the cyclotron frequency, that is, the magnetic field of the system, plays an essential role in Shannon's entropy. We observe that the more intense the magnetic field, the more localized is the Shannon probability density in the position space. Thus, we can conclude that the uncertainties in the particle localization measure will be less.

In contrast, in the momentum space, we have less localized probability densities when the field is intense. This fact leads us to conclude that the uncertainties in the measures related to the particle's momentum are more significant. Therefore, we conclude that with the results obtained from Shannon's entropy, we directly observe the validity of the Heisenberg uncertainty principle for the system studied.

With the energy spectrum found in the expression (\ref{energyM}) we use the Bose-Einstein statistic to describe the partition function and study the influence of the magnetic field on the statistical properties, namely, Helmholtz free energy, entropy, energy medium and specific heat. In all numerical calculations, we assume the parameters $m=\hbar=k_{B}=p_ {y} =\theta= 1$.

In fig. 9, we plot all profiles of thermal quantities as a function of temperature $\tau$ for different values of the magnetic field $\omega$, namely, $\omega=10,100$ e $1000$. From the thermodynamic properties, we notice that the Helmholtz free energy $\mathcal{F}(\tau)/N$ function decreases with temperature in an almost linear way. When the $\tau$ increases and for larger values of the magnetic field, conclude that the contribution of Helmholtz free energy becomes more significant. In other words, the free energy of Helmholtz tells us that for a stronger magnetic field it has a greater amount of energy to carry out the system's work. Therefore, the free energy to do work will be greater.

Meanwhile, when the magnetic field increases mean energy $\mathcal{U}(\tau)/N$ increases with an approximately linear behavior in the near of the origin, i. e., $\tau=0$. However, when $\tau$ increases, mean energy behavior starts to be more distinct. Clearly for values that the magnetic field is more intense, we will be in the high temperature regime and the energy assumed an exponential behavior. In this way, we conclude that for more intense fields the system has the capacity to increase the transfer of matter, or energy in the form of heat at the high temperature limit. Finally, we realized that for various values of the magnetic field, the heat capacity of the system tends to a constant, i. e., $\mathcal{C}_{v}(t\rightarrow\infty)\rightarrow k_{B}$. Therefore, for a given material the product of specific heat, at constant volume, by atomic mass or molecular mass is always constant.  This is the Dulong-Petit law.

\section{Conclusion}
In this work, we study the Shannon entropy of a spinless non-Hermitian particle in the presence of a magnetic field. Numerically, we conclude that Shannon's entropy satisfies the BBM relationship for ground states and excited states independent of the value of the magnetic field. We show how the thermodynamic properties changed as the magnetic field varies. Thus, we conclude that the magnetic field has an ability to modify the Shannon entropy which satisfies the BBM relationship of the model as well as the thermodynamic properties.

\section*{Acknowledgments} 
The authors thank the Conselho Nacional de Desenvolvimento Cient\'{\i}fico e Tecnol\'{o}gico (CNPq), grant n$\textsuperscript{\underline{\scriptsize o}}$ 308638/2015-8 (CASA), and Coordena\c{c}ao de Aperfei\c{c}oamento do Pessoal de N\'{\i}vel Superior (CAPES), for financial support. The authors also thank the anonymous referees for their valuable comments and suggestions.

\end{document}